\documentclass[preprint,12pt,authoryear]{elsarticle}

\usepackage{amssymb}
\usepackage{graphicx}
\usepackage{epstopdf, epsfig}
\usepackage{gensymb}
\usepackage{xcolor}
\usepackage{soul}

\journal{ArXiv}

\begin{document}

\begin{frontmatter}

\title{Kirigami Sheets in Fluid Flow}

\author[inst1]{A. G. Carleton}
\author[inst1]{Y. Modarres-Sadeghi\corref{mycorrespondingauthor}}
\cortext[mycorrespondingauthor]{Corresponding author}
\ead{modarres@engin.umass.edu }

\affiliation[inst1]{organization={Department of Mechanical Engineering},
            addressline={University of Massachusetts Amherst}, 
            city={Amherst},
            state={MA},
            postcode={01002}, 
            country={USA}}

\begin{abstract}
Kirigami patterned materials have found several applications in recent years due to their ability to assume complicated shapes and exhibit emergent physical properties when exposed to external forces. Consisting of an array of cuts in a thin material, fabrication of these patterns can be quite simple. Here we show that when they are placed in fluid flow, kirigami cut sheets with various patterns produce a verity of flow patterns in the wake. Through several sets of experiments, we show that the kirigami sheets placed in flow can undergo static or dynamic flow-induced instabilities as a result of which they can buckle or undergo limit cycle oscillations, or they can remain stable while undergoing very large elongations. These structural responses produce several different types of fluid patterns in the wake. We show that vortices both at small scales (scales comparable with the size of the individual kirigami cuts) and large scale (scales comparable with the size of the sheet) are formed in the wake of kirigami sheets. We also show that jets at different sizes can be formed and directed passively using kirigami sheets. These results show the potentials of kirigami sheets for passive flow control.
\end{abstract}

\end{frontmatter}



\section{Introduction}
Kirigami has recently received a lot of attention in several different fields of science and engineering including structural mechanics, robotics, and material science~\citep{blees2015graphene,choi2019programming,morikawa2018ultrastretchable,jin2020kirigami,rafsanjani2017buckling,zhang2015helical,sussman2015algorithmic,cheng2020kirigami,liu2021wallpaper,tang2019programmable,yang2021grasping}. This is due to the incredible ability of kirigami sheets to produce a myriad of complicated three-dimensional shapes through cutting certain patterns on an initially two-dimensional sheet and applying external forces on the sheet~\citep{konakovic2016beyond,callens2018flat,celli2018shape,li2018focused}. Despite the very large existing literature on the behavior of kirigami sheets in several different fields, their behavior in fluid flow is largely unexplored. Kirigami sheets in flow have wide-ranging potential applications in flow control.

\begin{figure}
\includegraphics[trim={0cm 0cm 0cm 0cm},width=1\linewidth]{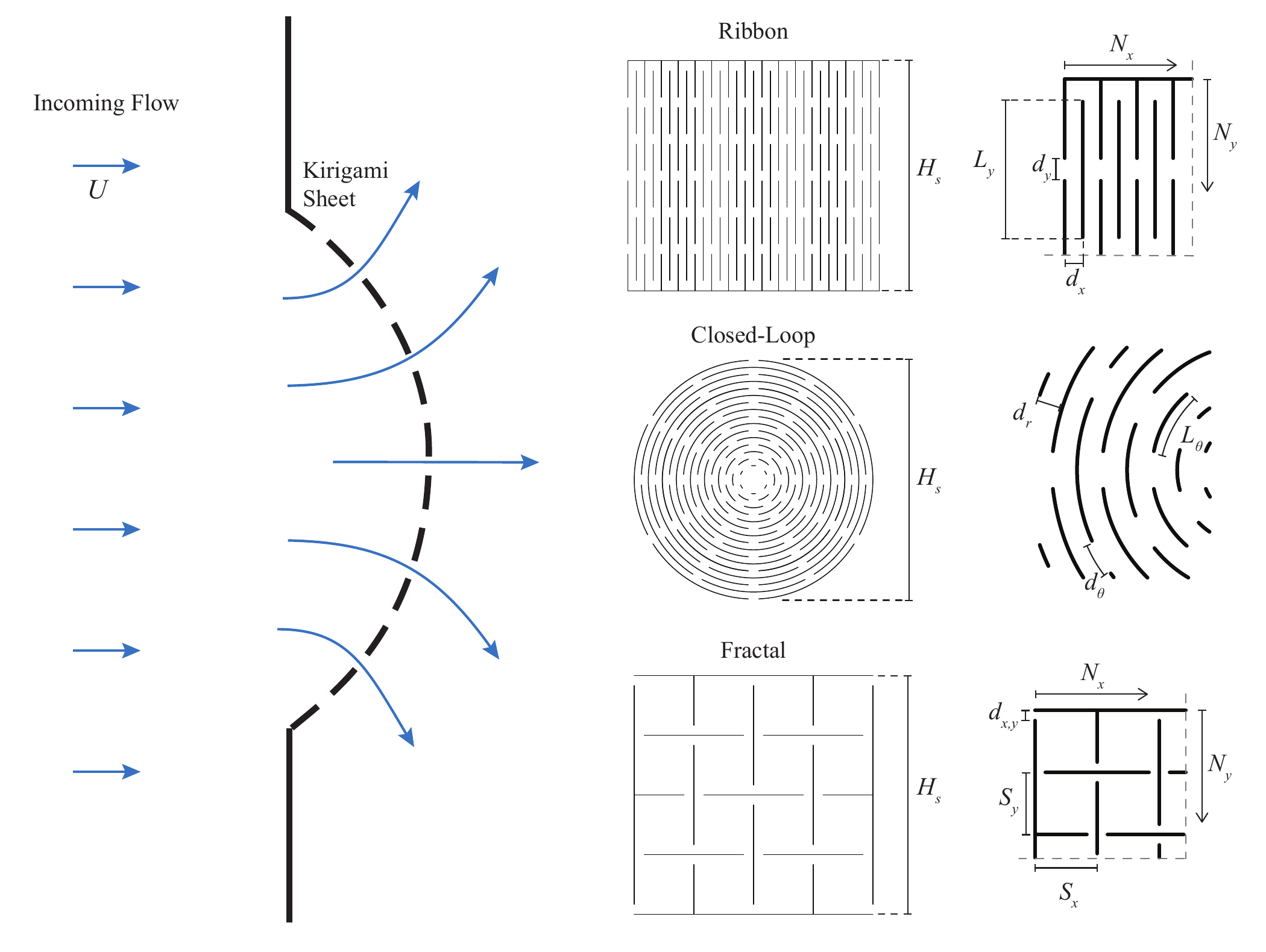}
 \caption{In this work, we investigate how a sheet with kirigami patterns behaves in flow and how it influences the fluid flow. We use kirigami designs with patterns that have been previously used for other applications, i.e., ribbon cut, close-Loop cut, and fractal cut patterns (in order from top to bottom). For the ribbon cut, $H_s=154$ mm, $d_y=7$ mm, $d_x=5.6$ mm, $L_y=42$ mm, and the number of cuts are $N_y=3$ and $N_x=15$. For the closed-loop cut, $H_s=216$ mm, $d_\theta=9.5$ mm, $d_r=6.4$, $L_\theta$ is dependent on the radius at that cut, and the number of cuts are $N_\theta=6$ and $N_r=16$. For the level-1 fractal cut, $H_s=152$ mm, $S_y=76$ mm, $S_x=76$ mm, $d_{x,y}=13$ mm, and the number of cuts are $N_y=2$, and $N_x=2$. For the level-2 fractal cut, $H_s=152$ mm, $S_y=38$ mm, $S_x=38$ mm, $d_{x,y}=6.4$ mm, and the number of cuts are $N_y=4$ and $N_x=4$.}
    \label{fig:schematic}
\end{figure}

Kirigami designs consist of a pattern of slits, or a combination of slits and creases,
which makes them different from origami designs that are solely based on creases. The ease of production and deployment (which is often done simply by applying external tension on a 2D sheet) has caused several recent applications for kirigami sheets in ``structural mechanics, materials, optics, electronics, robotics, and bioengineering"~\citep{sun2021geometric}. This recent review article covers tens of references for kirigami applications, which we do not repeat here. Despite this extensive use in other fields, kirigami designs have not been used for fluid mechanic applications, except for very limited number of studies as we will discuss later.

The kirigami designs are either cut-only or cut-and-fold kirigami patterns. The cut-only patterns are categorized into three general groups of (i) ribbon-cut kirigami, (ii) closed-loop kirigami, and (iii) fractal-cut kirigami~\citep{sun2021geometric}. Samples of these categories together with the parameters that define them are shown in Figure~\ref{fig:schematic}. All these kirigami cuts can achieve very large strains---by far more than the strain that could have been achieved by the original uncut sheet. An early proposed application of \textit{ribbon cut} kirigami patterns was to improve the ductility of graphene~\citep{blees2015graphene,grosso2015bending}, and later on for several other applications in shape memory alloys, metallic glasses and carbon nanotubes~\citep{an2020modeling,han2021critical,zhao2017carbon}. When a kirigami sheet with parallel cut patterns is stretched, the sheet displacement changes nonlinearly with increasing external tension~\citep{isobe2016initial,Huang13,tang2017programmable}. Initially the sheet has a relatively high stiffness because as the slits begin to expand, a bending moment is created perpendicular to the sheet's surface, and the individual strips have a higher moment of inertia in this direction. As the displacement of the sheet increases, the strips begin to rotate out of plane, bringing the local orientation of the sheet surface closer to parallel with the bending moment created by the opening slits. In this orientation, the moment of inertia is significantly less, due to the large strip-width to sheet-thickness ratio, and the effective stiffness of the sheet decreases. Then, as the displacement is further increased, the slits approach the limit of how far they can open and the stiffness of the sheet begins to increase. 
\textit{The closed-loop} kirigami patterns were recently  investigated for nano-scale materials applications~\citep{li2018focused,liu2018nano}. The main characteristic of these designs is that the parts of the sheet within a closed loop pattern influence each other when they are under tension, and the final 3D shape depends both on the cuts and the geometric topology and stress equilibrium of the closed-loop~\citep{sun2021geometric}.
\textit{The fractal cut} kirigami patterns are used mainly to create rotating units within the sheet when external tension is applied. In fractal cuts, the same cut patterns is repeated for each unit, resulting in cuts with finer details and larger strains as the levels are increased. Strains up to \%62~\citep{cho2014engineering} are reported for square fractal cuts (shown in Figures~\ref{fig:schematic}) and much higher values of \%156~\citep{tang2017design} for rectangular fractal cuts.

In some recent studies, kirigami patterns have been deployed in fluid flow, but in contexts different from what we discuss here. In one of these studies~\citep{gamble2020multifunctional}, an airfoil was covered with a kirigami ``skin'', which was then tensioned externally, creating a three-dimensional geometry that increased drag, enabled yaw control, and delayed aerodynamic stall. In another study~\citep{li2021aerodynamics}, a rigid three-dimensional kirigami-based geometry was used to control the wind flow and form counter-rotating vortices such that droplets could be ejected from the incoming fog clusters. In both of these studies, the kirigami design was actuated through an external force to influence the flow, and it was not actuated due to the flow forces. In a recent study, \cite{marzin2022flow} investigated how a sheet with a simple kirigami pattern (ribbon cuts) reacts to the incoming flow, by quantifying the sheet's static deflection versus increasing flow velocity. They showed that the sheet's  deflection can be controlled by changing the number and size of the cuts. They also showed that both the shape of the pore and the slit arrangement influence the elasticity of the kirigami sheet, which in turn influences the amount of the sheet deflection. Note that while the kirigami sheet in their case was placed in water flow, the main focus was on the elasticity of the structure and not the flow behavior as a result of its interaction with the flow.
Additionally, the flow was allowed to pass around the sides of the kirigami sheet in this case, and not forced through the pattern.

The flow forces that act on the bent kirigami sheet will have two components, in the normal, $F_{N}$, and tangential, $F_{T}$, directions of the local curvature of the sheet~\citep{marzin2022flow}:

\begin{equation} \label{eq:1}
F_{N/T}=\frac{1}{2} C_{N/T}(\epsilon) H (a(\epsilon) \mathbf{U.n})^2,
\end{equation}

\noindent
where $C_{N/T}$ are the flow force coefficients in those directions, $H$ is a characteristic length of the sheet, $\mathbf{U}$ is the incoming flow velocity, $\mathbf{n}$ represents the normal direction, $a$ is a factor to consider the blockage effect (the sheet sees a flow velocity of $aU$), and $\epsilon$ is the elongation of the sheet. It is expected that the flow forces that act on a kirigami sheet change with the elongation of the sheet. An essential point that should be considered here is that similar to any other 
Fluid-Structure Interaction (FSI) 
problem, $\epsilon$ does not necessarily stay constant when the flow acts on a kirigami sheet. In fact in this particular problem, even if the elongation stays constant, flow forces do not, due to the interactions between flow that goes through the cuts and the structure around it at a small scale, as well as the interactions between flow in the wake of the structure with the deployed sheet at the large scale of the sheet. The flow forces that act on the sheet at the cut scale are proportional to the size of the cuts and use a characteristic length scale of $H=H_c$ (cut scale) in Equation (1), and together, they produce forces that scale with $H=H_s$ (sheet scale) that influence the overall morphing of the sheet. 

Here, we focus on how the fluid flow is affected by kirigami sheets that are placed perpendicular to  the flow direction. The question is how the slits that are cut in different kirigami patterns affect the wake of the sheet when the sheet is placed in a uniform cross flow. In our experiments, we block the incoming flow with an initially two-dimensional kirigami sheet such that the fluid has no other path except through the kirigami patterns. The goal is to exhibit the ability of relatively simple kirigami patterns to produce a variety of complicated flow patterns in the wake of the sheet. We consider three distinct types of kirigami designs: ribbon cuts, closed cuts and fractal cuts.

\section{Experimental Setup}
The sample kirigami patterns were designed with CAD software and then cut out of 0.1 mm thick polyester film. This material has a tensile strength of 190 MPa and a flexural modulus of 4.8 GPa. These cut sheets were then clamped around their perimeter to a reducing nozzle
fixture that was placed in the test section of a re-circulating water tunnel with a test-section of 50 cm $\times$ 38 cm $\times$ 127 cm, with a uniform flow profile~\citep{carleton2022passive}. The turbulence intensity, defined as $TI=u'/U$, where $U$ is the mean flow velocity and $u'$ is the standard deviation of flow velocity, is calculated to be around 1\% or less directly upstream of the sheet. The maximum flow velocity tested here was $U_{max}=13$ cm/s for all cases. A Phantom Miro M110 high speed camera ($1280\times 800$ pixels operating at 100 fps) was used to record the response of the structure. The high contrast of the recordings allowed algorithmic edge detection of the kirigami sheet. We used an in-house image analyzing code that batch processes images with morphological tools and can generate time-series data for the points along the trailing edge of the kirigami sheet. These data were used to determine the characteristic deployment shape and magnitude of each pattern as a function of flow velocity and time when oscillations occurred. 

Flow visualization was conducted using the Bubble Image Velocimetry (BIV) technique that we have extensively used in the past for wake visualization in other systems~\citep{currier2021dynamics,carlson2021flow,carleton2022passive}. In this method, a 50 $\mu$m stainless steel wire was mounted at the midplane of the test section, crossing 5 cm upstream from the kirigami sheet. A 100 volt, 2 amp power supply connected the stainless steel wire to carbon plates below the wire, and the voltage potential between the stainless steel and carbon created a stream of hydrogen bubbles of similar scale to the wire diameter via hydrolysis. These bubbles were small enough that the buoyancy forces were negligible compared to the drag forces and the bubbles therefore followed the local flow velocity. The bubble field was illuminated by two light-emitting diode (LED) lighting banks for imaging by 
the Phantom camera. Additionally, we generated tight fitting masking files from the tracked data to improve the detailed observable in the BIV analysis. The wake was quantified using PIVlab, a MATLAB toolbox for performing particle image velocimetry~\citep{Thielicke_2021}. The bubbles were used as the tracer particles, and a velocity vector field, $\mathbf{u}$, was generated that represented the local flow velocities by implementing a cross-correlation algorithm on successive image pairs. We then calculated the vorticity vector field using this velocity field by taking the curl of the velocity field as $\mathbf{\omega}=\nabla\times\mathbf{u}$. This vorticity field is plotted in the wake visualization figures that will follow.

\begin{figure}
    
  \centering
  \includegraphics[trim={0cm 0cm 0cm 0cm},width=1\linewidth]{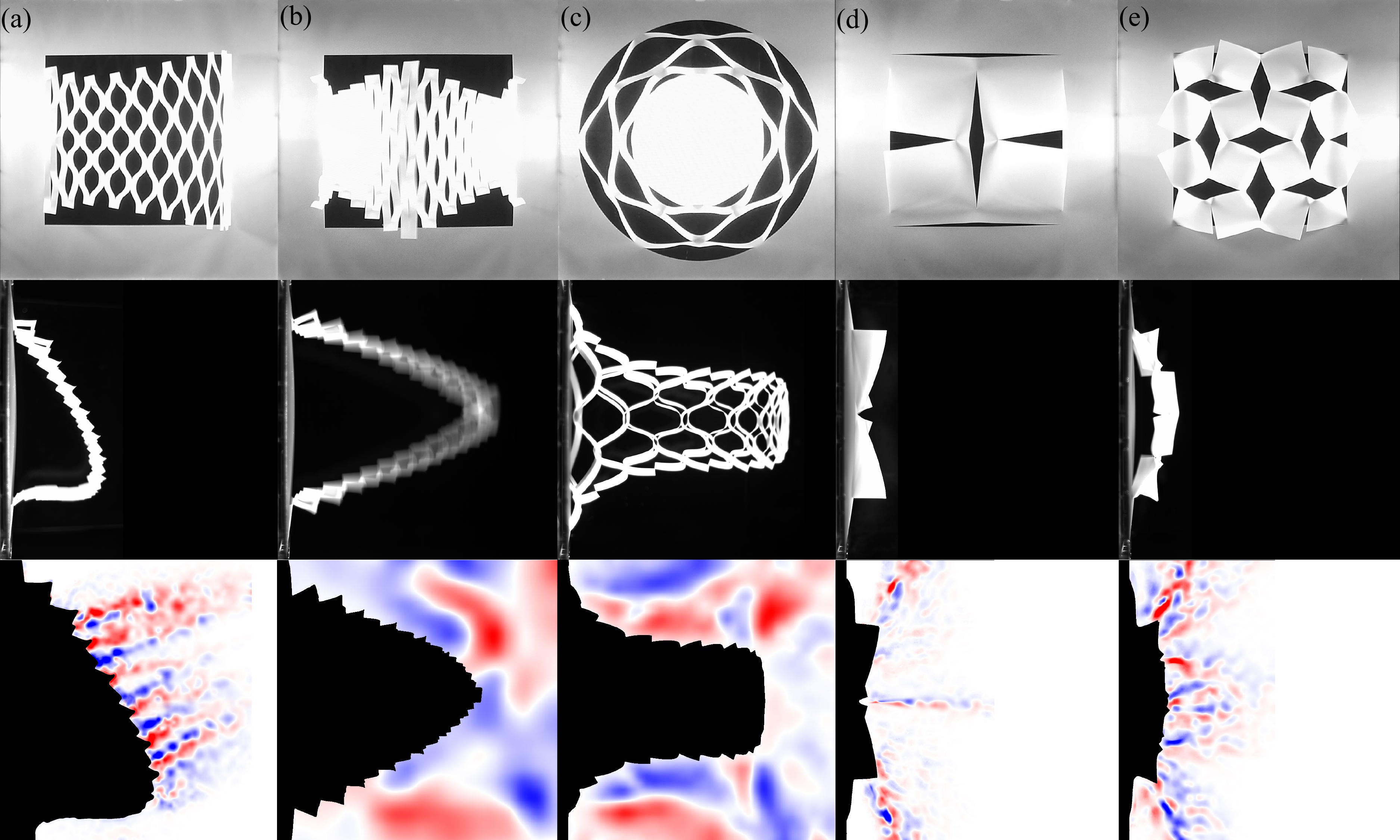}
 \caption{Snapshots of kirigami patterns deployed in fluid flow. Upper row is a view from downstream (flow comes out of the plane); the middle row is a side view (flow is from left to right), and the lower row corresponds to the observed wake along a horizontal plane at the midsection of the sheet. The first two cases, (a) and (b), are ribbon cuts, (a) when the strips are all biased to one side and (b) when the strips are biased symmetrically toward the center of the sheet. The third case (c) is a closed-curve kirigami pattern. Cases (d) and (e) correspond to fractal kirigami cuts with square units at level-1 (d) and level-2 (e). The snapshot of the middle row for case (b) is a long exposure to illustrate the oscillatory response of that case. Videos of the kirigami responses as the flow velocity is increased for all five cases are provided as supplementary materials.
 }
 
\label{fig:prel_DownStreamAndSideViews}
\end{figure}

\section{Results} 
We consider three kirigami designs from distinctly different patterns that have been used in other applications~\citep{sun2021geometric}. These patterns are shown in Figure~\ref{fig:prel_DownStreamAndSideViews}. The first two columns from the left ((a) and (b)) correspond to a ribbon cut (or parallel cut) pattern. Once a sheet is cut with this pattern, the resulting strips can be manually biased sideways (a) or toward the center (b). The third column (c) corresponds to a closed-loop cut kirigami pattern, where a symmetric pattern evolves around its center and the cuts stay within closed loops. The last two columns, (d) and (e), correspond to fractal cut kirigami patterns, in which a cut pattern is used for level-1 (d) and is scaled and repeated in level-2 (e). 

\subsection{Multi-scale vortex shedding and flow-induced vibrations of ribbon-cut kirigami sheets} 
A strategy to enhance mixing in fluid flow is to increase turbulence and generate flow unsteadiness, which is achieved by forming vortices in an incoming uniform flow~\citep{gad2003flow}. 
Here, we show how vortices at multiple scales (i.e., the scale of the cuts, $H_c$, as well as the scale of the sheet, $H_s$) can be produced by using sheets with ribbon-cut kirigami patterns.
Figures~\ref{fig:prel_DownStreamAndSideViews}(a) and (b) correspond to the behavior of a ribbon-cut kirigami pattern, when placed in a fluid flow. In this type of pattern, an array of parallel slits that are offset from each other are cut in the sheet so that they create an expanded mesh shape when deployed. As each slit expands, the strips of the sheet on either side of it bend and rotate, and act as small-scale fins. When placed in flow, these small fins interact with the fluid that goes through the slits, and as a result of this interaction forces from the fluid act on the strips. These forces that act at the scale of strips (and scale as $H_c$ in Equation (1)), then collectively generate relatively large fluid forces (that scale as $H_s$) on the sheet, which cause the response of the sheet at large scale. As we will see later, the large-scale response of the sheet then results in changes in the relative orientation of the strips with respect to the incoming flow, and changes the small-scale flow forces that act on the strips, which in turn change the total force acting on the structure, thereby constituting a multi-scale FSI system. 

Once the slits are cut in a ribbon cut kirigami pattern, and before placing them in fluid flow, the orientations of the strips can be biased manually to impose a desired response on the large sheet. Here, we consider two configurations for the biased strips: first, we bias all strips in the same direction to produce an inherently asymmetric structure (Figure~\ref{fig:prel_DownStreamAndSideViews}(a)), and then we bias them such that the strips on the two sides of the sheet are rotated toward the center of the sheet, which result in an inherently symmetric structure (Figure~\ref{fig:prel_DownStreamAndSideViews}(b)).

\begin{figure*}
\centering
  \includegraphics[trim={0cm 0cm 0cm 0cm},width=1\linewidth]{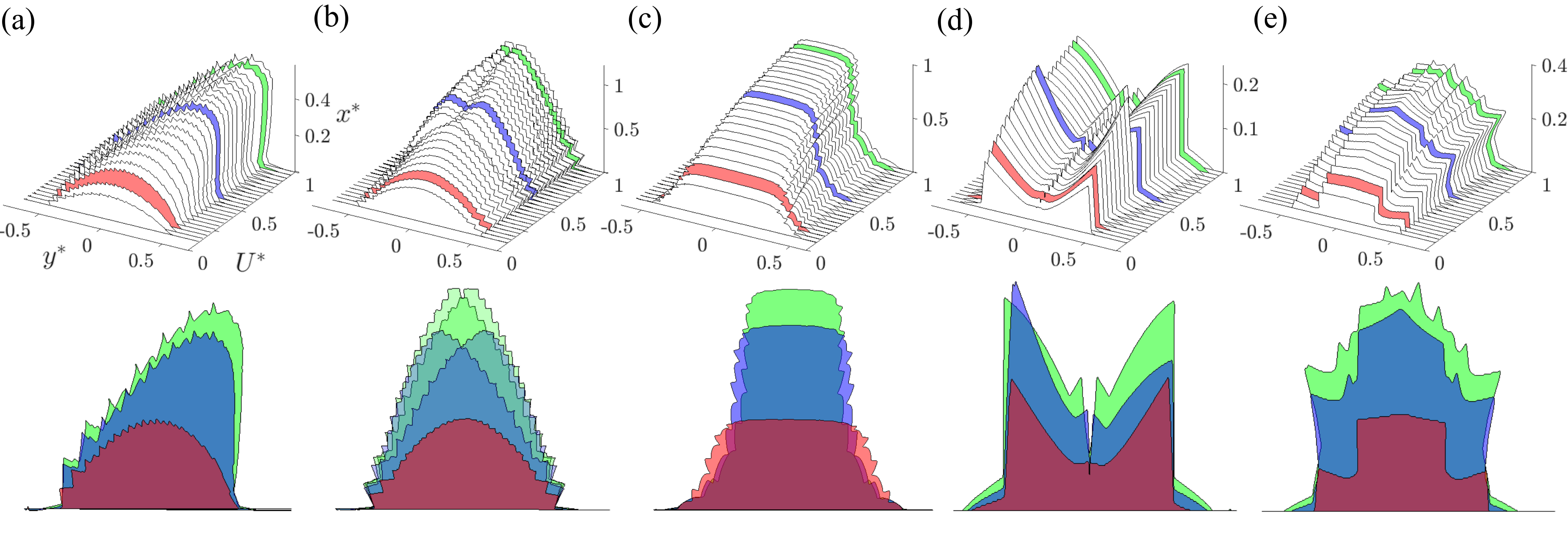}
 \caption{Tracked displacements of the outer edge of the kirigami sheet ($x^*=x/H_s$, and $y^*=y/H_s$, where $x$ is the displacement in the direction of flow, $y$ is perpendicular to the flow, and $H_s$ is the characteristic sheet scale) versus normalized flow velocity, $U^*=U/U_{max}$. (a) A ribbon-cut pattern when the strips are biased in one direction, (b) a ribbon-cut pattern when the strips are biased toward the center, (c) a closed-curved pattern, (d) a level-1 fractal pattern, and (e) a level-2 fractal pattern. Displacements in the $x^*$ direction are exaggerated for better visibility of the details. For the symmetrically biased ribbon cut shown in column (b), the oscillations are demonstrated by showing the outline of the sheet at both cross flow maxima.} 
\label{fig:TrackedDisplacements}
\end{figure*}

At the instant shown in the snapshot of Figure~\ref{fig:prel_DownStreamAndSideViews}(a) (for an asymmetrically biased sheet), the sheet is bent sideways as was also observed before for such kirigami patterns~\citep{marzin2022flow}. This large-scale ($H_s$) deflection of the sheet has resulted in the formation of small-scale ($H_c$) openings and small-scale ($H_c$) fins in between these openings. Each small-scale fin then acts as a ``bluff'' body (as opposed to a streamlined body that would disturb the incoming flow minimally) that is placed in fluid flow, and it is expected that vortices are formed around such bodies. This vortex shedding is indeed observed through the BIV calculations of vorticity in the wake of the structure in the lowest panel of Figure~\ref{fig:prel_DownStreamAndSideViews}(a), where blue corresponds to vortices that rotate in the clockwise direction and red to the vortices that rotate in the counterclockwise direction. The blue vortices are shed from the top of each small-scale fin and the red ones from the bottom. The size of these vortices scales with  $H_c$, and since there are several small-scale fins in the side of the structure from which the vortices are shed (6 in the case shown in the figure), several (6 in this case) sets of these vortices are observed in the wake in parallel. These shed vortices cause a dramatic increase in turbulence downstream of the kirigami sheet, with the turbulence intensity increasing from 1\% or less upstream of the sheet to 28\% or more in the region of flow exiting the sheet, depending on the incoming flow velocity and specific cut parameters used. Turbulence intensity values were calculated from the BIV results, following the method used by~\cite{gomes-fernandes_ganapathisubramani_vassilicos_2012}, and the values obtained for flow in the wake of the sheet are on par with the higher end of traditional grid-induced turbulence that is previously observed~\citep{Takahiro_KIWATA20142014jfst0056}. Unlike standard methods of inducing grid turbulence, however, kirigami sheet induced turbulence varies with flow velocity, due to the decreased blockage ratio that occurs as the sheet expands. Changes in this value have a significant effect on the downstream turbulence intensity~\citep{LIU2004307}, and therefore kirigami patterns could potentially be designed such that the downstream turbulence follows a desired trend with respect to incoming flow velocity. 

The orientation that the shed vortices travel downstream is dictated by the orientation of the deflected sheet, which is dictated by the flow forces at the sheet scale, $H_s$. The progression of the large-scale deflection of the sheet as the flow velocity is increased is shown in Figure~\ref{fig:TrackedDisplacements}(a). The sheet's deflection increases with increasing flow velocity, since flow forces that act on the structure increase as $\mathbf{U}^2$ (Equation (1)). For small values of flow velocity, the sheet expands nearly symmetrically in the direction of the incoming flow (Figure~\ref{fig:TrackedDisplacements}(a)), since the flow forces that act tangentially on the strips ($F_T$) are small at low flow velocities and cannot exert the necessary force on the sheet in that direction to impose large-scale asymmetry. Note that these forces scale with $H_c$ and at small values of $\textbf{U}$, their magnitude remains small.  As higher flow velocities (larger $\textbf{U}$), the tangential forces, $F_T$, that act on the strips increase and since the strips in this case are biased in one direction, collectively, impose a large sideway force on the structure, which then imposes a noticeable asymmetry on the structure. The bending of the sheet increases with increasing flow velocity, which implies that the orientation of where the shed vortices travel downstream also changes with increasing flow velocity. The sheet stays static for the entire range of flow velocities tested here. The stiffness of these sheets change if the number and size of the slits are changed~\citep{marzin2022flow}, which would mean a different direction for the shed vortices in the wake. It is also clear from the visualisation results of Figure~\ref{fig:prel_DownStreamAndSideViews}(a) that a change in the number of slits would change the number of vortices that are shed in the wake.

\begin{figure}
  \centering
  \includegraphics[trim={0cm 0cm 0cm 0cm},width=.8\linewidth]{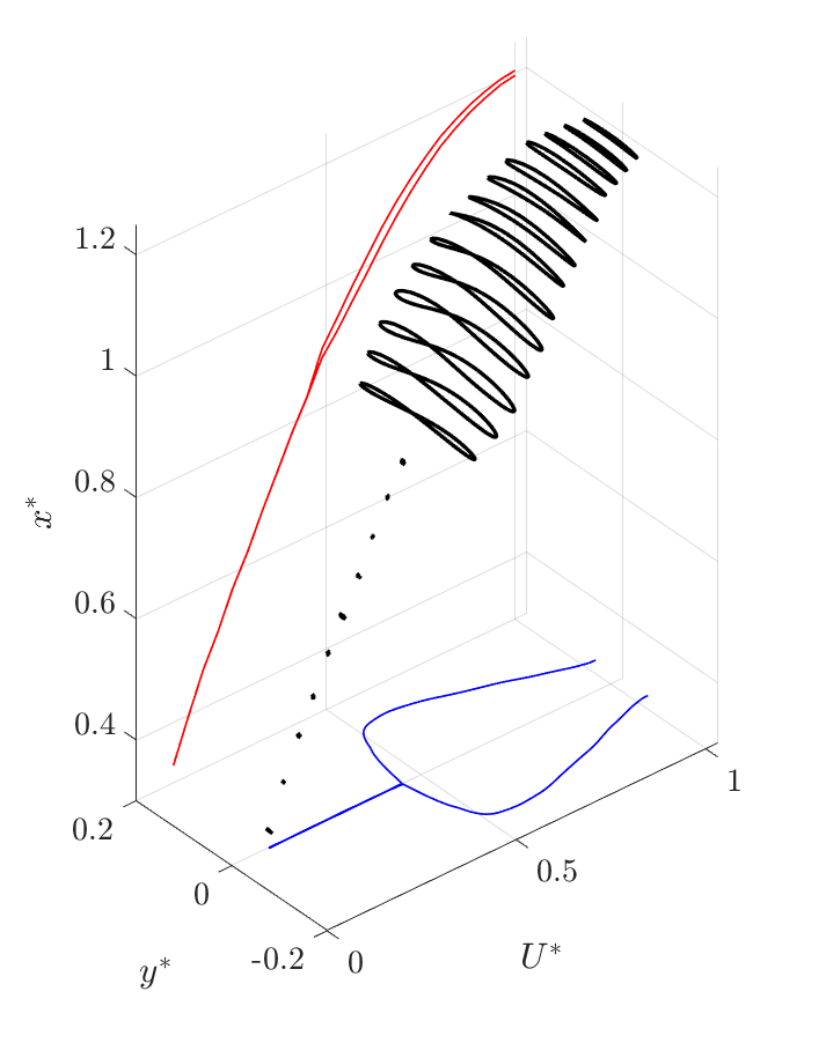}
 \caption {The tip displacements of the kirigami sheet with ribbon-cut pattern and strips biased symmetrically toward the center  (Figure~\ref{fig:prel_DownStreamAndSideViews}(b)). These are normalized displacements in the direction of flow, $x^*=x/H_s$, 
 and perpendicular to that, $y^*=y/H_s$, as well as the observed trajectory of the tip versus normalized flow velocity, $U^*=U/U_{max}$. The sheet elongates symmetrically for low flow velocities. Then at $U^* \approx 0.4$, it undergoes a dynamic instability, which leads to limit cycle oscillations of the sheet. The tip trajectories when limit cycle oscillations start follow a figure-eight pattern initially, and then a teardrop pattern at higher flow velocities. The red and blue lines show the bounds of motion in the $x$ and $y$ directions, respectively.}
    \label{fig:P18TrackedMotion}
\end{figure}

In the snapshots of Figure~\ref{fig:prel_DownStreamAndSideViews}(b) (for a symmetrically biased sheet), the sheet undergoes limit cycle oscillations. When oscillations are observed, the flow pattern in the wake of the structure is entirely different from that in the wake of a static structure (compare Figures~\ref{fig:prel_DownStreamAndSideViews}(a) and (b)). Instead of small-scale ($H_c$) shedding patterns observed in the wake of a static sheet, when the sheet oscillates, large-scale ($H_s$) vortices are shed at a frequency equal to the sheet's oscillation frequency. The fluid flow that approaches the sheet goes through the slits and small-scale vortices are formed during each half-cycle of oscillations, similar to the pattern observed in Figure~\ref{fig:prel_DownStreamAndSideViews}(a). However, in this case, when the sheet stops at the end of each half cycle of its oscillations, the small-scale vortices are released simultaneously and form the large-scale vortices observed in Figure~\ref{fig:prel_DownStreamAndSideViews}(b). Initially the sheet elongates statically in the direction of flow, and at a critical flow velocity, it develops limit cycle oscillations. This transition to oscillations is marked by the appearance of two peaks in Figure~\ref{fig:TrackedDisplacements}(b) that correspond to the two extreme positions of the sheet during its oscillations. A summary of the sheet's response in this case is given in Figure~\ref{fig:P18TrackedMotion} by plotting the normalized displacements of the tip of the sheet in the direction of flow, $x^*=x/H_s$, and perpendicular to that, $y^*=y/H_s$, and in the form of its tip trajectory. It is clear that the sheet initially expands symmetrically in the direction of flow, $x^*$. Then oscillations start at a normalized flow velocity of $U^* =U/U_{max} \approx 0.4$. Once the oscillations start, the tip of the sheet initially follows a figure-eight trajectory, suggesting a 2:1 ratio between oscillation frequencies in the $x$ and $y$ directions. The figure-eight trajectories grow in their size initially, reaching a maximum that is clearly visible in the $y^*$ plot, and then decrease in magnitude as the mean displacement of the sheet in the direction of flow increases. At higher flow velocities, the trajectories become teardrops, suggesting that the symmetry of the response breaks in this case as well.

The differences in the responses of the sheet in this case and the previous case are a result of the orientation of the fins. When the fins are biased asymmetrically, the small-scale forces that act on the fins result in a collective force that bends the sheet sideways. In the case where the fins are biased symmetrically toward the center, the small-scale flow forces result in a collective large force that changes its orientation in each half cycle and causes limit-cycle oscillations of the sheet. 

\subsection{Large-scale rotational flow without large-scale structural instability using closed-curve kirigami sheets} 

The large-scale rotational flow that led to large-scale mixing we observed in the previous section was a result of the large-scale flow-induced instability of the structure. This might give the impression that large-scale rotational flow using kirigami sheets relies on large-scale instability of the structure. Here, by using a different kirigami pattern, we show that large-scale rotational flow can be obtained even for cases where the structure stays stable. The plots in Figure~\ref{fig:prel_DownStreamAndSideViews}(c) correspond to the response of a closed-curve kirigami pattern in flow. In this pattern, curved cuts are made at different distances from the center of a circle. Naturally, the length of the curved cuts increases with the distance of the cuts from the center. The symmetric nature of these cuts makes it possible for the middle disc to stay parallel to its original orientation (i.e., perpendicular to the flow) as the flow velocity is increased. The snapshot of Figure~\ref{fig:prel_DownStreamAndSideViews}(c) corresponds to a case where the kirigami sheet has undergone a large elongation in the direction of flow, but despite this large elongation, the sheet has not experienced any flow-induced instability. The inherent design of this pattern stops the fluid from continuing to flow in its original direction, due to the presence of the middle disc that stays perpendicular to the flow at all times. Instead, the flow is deflected to the sides and goes through the openings of the sheet. At lower flow velocities when the sheet is only slightly stretched, parallel vortices are shed from the sides of the sheet at a scale of $H_c$, similar to those shed from the sides of the deflected ribbon cuts in Figure~\ref{fig:prel_DownStreamAndSideViews}(a). As the sheet is stretched more, the slits expand and form large and well-structured openings on the sides of the sheet (Figure~\ref{fig:prel_DownStreamAndSideViews}(c)). As a result, the flow goes through the openings with minimal interactions with the strips and forms regions of rotating flow on the two sides of the structure at a scale of $H_s$ (Figure~\ref{fig:prel_DownStreamAndSideViews}(c)). Figure~\ref{fig:TrackedDisplacements}(c) shows that the sheet elongates smoothly as the flow velocity is increased and no large-scale structural instability is observed. For all these flow velocities, almost still fluid is observed behind the central disc of the sheet, despite the large incoming flow velocity, since the fluid flow is always directed to the sides. Since the structure does not oscillate in this case (as opposed to the case of Figure~\ref{fig:prel_DownStreamAndSideViews}(b)), the fluid rotates in opposite directions to those around the structure in Figure~\ref{fig:prel_DownStreamAndSideViews}(b). The red vortex is immediately on the top of the sheet in Figure~\ref{fig:prel_DownStreamAndSideViews}(c) and the blue vortex is below, and vice versa in Figure~\ref{fig:prel_DownStreamAndSideViews}(b). While in both cases large-scale vortices are observed in the wake of the sheet, these vortices are formed differently. The formation of large-scale rotating flow in the closed-curve pattern of Figure~\ref{fig:prel_DownStreamAndSideViews}(c) is due to the bending of the incoming flow that cannot go through the end-disc, but in the ribbon-cut pattern of Figure~\ref{fig:prel_DownStreamAndSideViews}(b), the large-scale rotational flow is formed as a result of the sudden change in the direction of motion of the oscillating structure.

\subsection{Jet formation using fractal-cut kirigami sheets}

Here, we show that jets can be passively formed and directed at predictable angles using kirigami sheets. Figures~\ref{fig:prel_DownStreamAndSideViews}(d) and (e) show snapshots of the response of kirigami sheets with square fractal cuts at level-1 (d) and level-2 (e), when placed in fluid flow. The fractal cut kirigami patterns are used mainly to create rotating units within the sheet when external tension is applied. In fractal cuts, the same cut patterns is repeated for each unit, resulting in cuts with finer details and larger strains as the levels are increased~\citep{cho2014engineering,tang2017design}. The level-1 cuts of Figure~\ref{fig:prel_DownStreamAndSideViews}(d) divide the large square to four smaller squares, and the level-2 cuts (Figure~\ref{fig:prel_DownStreamAndSideViews}(e)) repeat these cuts in each unit of level-1, resulting in 16 smaller squares in the level-2 sheet. In the snapshots of the figure, the sheets have undergone flow-induced buckling, which has resulted in the formation of well-structured openings in the sheets. Fluid flows through these openings and forms distinct jets in the wake at a size comparable with $H_c$ as shown in the flow visualization plots of Figures~\ref{fig:prel_DownStreamAndSideViews}(d) and (e). The number of openings in each buckled case dictates the number of jets that are formed in the wake, and as a result, 3 jets are observed in the wake of the level-1 sheet (one opening in the middle and two on the sides) and 5 in the wake of level-2.

\begin{figure}
  \centering
  \includegraphics[trim={0cm 0cm 0cm 0cm},width=.8\linewidth]{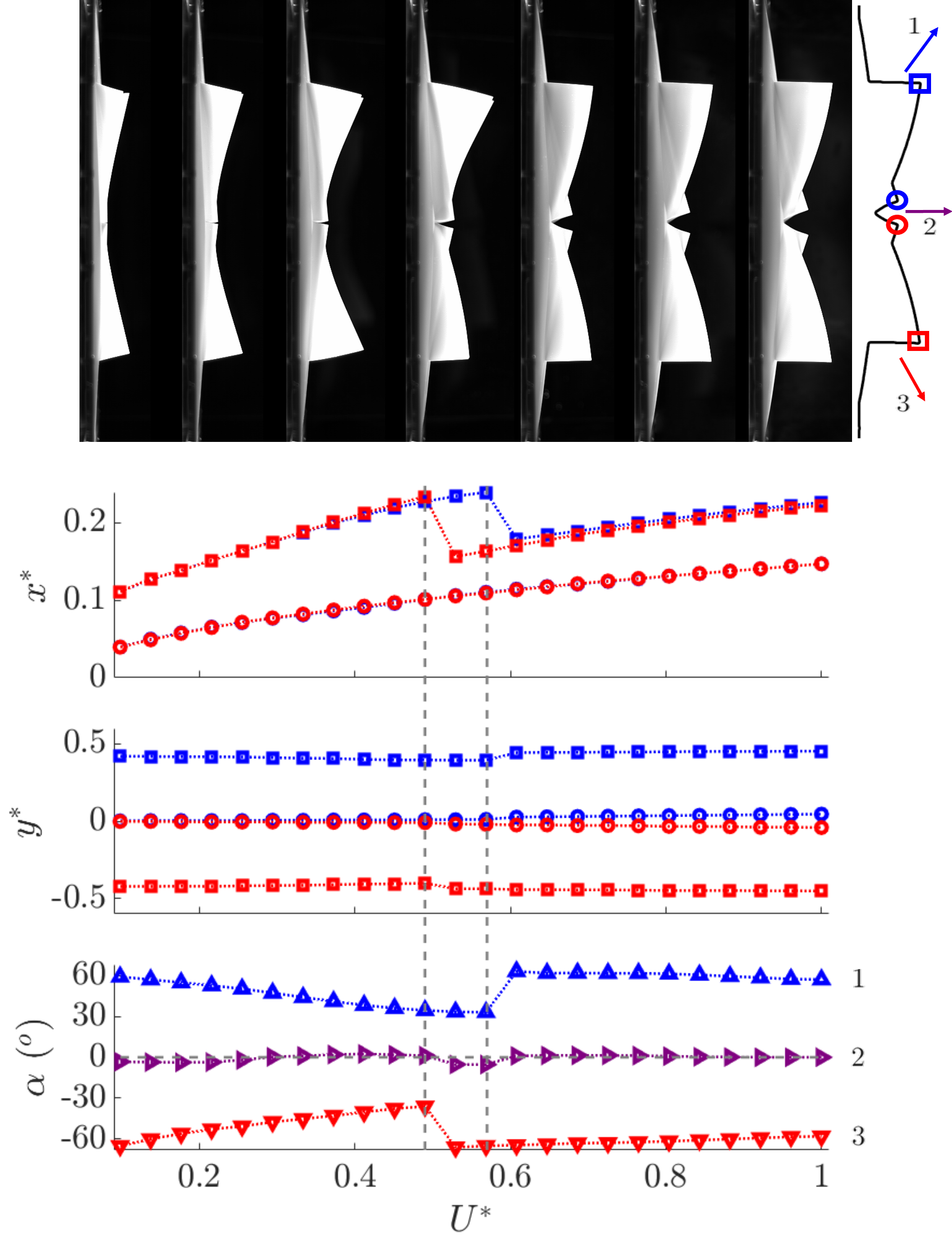}
  \caption {The response of the kirigami sheet with level-1 fractal cut pattern as normalized flow velocity, $U^*$, is changed. Side view snapshots from the experiments, tracked displacements of four points on the kirigami sheet in the direction of flow, $x^*$, and perpendicular to that, $y^*$, together with the directions of jets in the wake. The onsets of the two buckling instabilities are highlighted by vertical dashed lines. The angles of the jets are assumed to be positive counterclockwise with $0^{\circ}$ defined to be in the direction of the incoming (left to right).}
    \label{fig:FractalTipTracking}
\end{figure}

The directions of the jets in the wake of the sheet are dictated by the state of the structure, which itself depends on the flow velocity, as shown in Figure~\ref{fig:FractalTipTracking}. For the level-1 sheet, at low flow velocities, the two side jets are directed at angles of $\pm 60^{\circ}$. The middle jet is weak, since the middle slit of the sheet is not open yet, and is shed at $0^{\circ}$. As the flow velocity is increased, the two sharp edges of the sheet (denoted by square symbols in Figure~\ref{fig:FractalTipTracking}) are influenced by the mean drag force (which scales with $H_s$) and move continuously in the flow  direction, which then causes the two side jets to be 
redirected at smaller angles with respect to the flow direction. The angles of the side jets decrease continuously and reach around $30^{\circ}$ as the flow velocity is increased to $U^* \approx 0.5$ (Figure~\ref{fig:FractalTipTracking}). At this point, the first structural instability occurs and the lower side of the structure buckles (the lower edge, denoted by a red square, moves to the left in the snapshot). This is observed in the form of a sudden drop of the $x^*$ displacement and a slight increase in the magnitude of $y^*$ in Figure~\ref{fig:FractalTipTracking}. Naturally this buckling changes the sizes of the middle and lower openings and causes an asymmetry in the structure, therefore influencing the direction of the jets that are formed through those openings. After this structural instability, Jet 3 goes back to an angle of $-60^{\circ}$, and Jet 2, that was initially oriented horizontally, deflects downward slightly. Jet 1 follows the original trend and its angle continues to decrease. This behavior of Jet 1 changes as soon as the second instability of the structure occurs and the upper side of the sheet buckles (the upper edge, denoted by a blue square, moves to the left in the snapshots). This is the mirror image of the first instability, and the structure becomes symmetric again. As a result, Jet 1 follows a behavior similar to what Jet 3 followed after the first structural instability, and its orientation goes back to $+60^{\circ}$. After $U^* \approx 0.6$, the structure remains stable, exhibiting only a modest amount of further elongation as the flow velocity is increased. This elongation in turn causes a slight reduction in the angle of the side jets, while the middle jet stays at $0^{\circ}$, since the structure remains symmetric.

The five jets that are observed in the wake of the sheet with a level-2 cut also change their directions as the flow velocity is increased, as dictated by the shape of the sheet at each flow velocity. At lower flow velocities, two jets (1 and 2) are directed at $+70^{\circ}$ and $+60^{\circ}$, one jet (3) is directed at 0 and two more (4 and 5) are directed at $-60^{\circ}$ and $-70^{\circ}$. As the flow velocity is increased, jets 2 and 4 reorient themselves from $\pm50^{\circ}$ to $0^{\circ}$ very quickly by $U^*\approx0.2$, and they stay at that angle for all higher flow velocities, since the sizes of the openings through which these jets are formed stay constant. Jets 1 and 5 also change their directions as the flow velocity is increased, albeit with a much slower rate, and their angles decrease from a magnitude of $70^{\circ}$ linearly to $50^{\circ}$.
In the visualization image of Figure~\ref{fig:prel_DownStreamAndSideViews}(d), the side jets are wider and the middle jet is more focused. These sizes follow the sizes of openings through which these jets are formed. The plot of the figure is a view from the top of a horizontal plane at the midsection of the sheet, where the openings are wider on the sides and smaller in the center. If observed from a vertical plane, the jet in the middle comes out of a wide opening and is diffused in that orientation, and the jets on the side come from much smaller openings, and as a result they are much more focused. The sizes of jets observed at the midsection of the level-2 kirigami cuts of Figure~\ref{fig:prel_DownStreamAndSideViews}(e) follow the sizes of the opening at the midsection as well. The jet in the middle and the two extreme side jets are the wider jets and the jets that are in between are more narrow when observed along a horizontal plane, and opposite when observed along a vertical plane. 

\section{Conclusions}
When a kirigami sheet is placed perpendicular to the incoming fluid flow, the deformation of the sheet that is needed to deploy the kirigami pattern is passively caused by the fluid forces, without any need to apply external tension. This enables the use of kirigami sheets as completely passive flow control devices, which can be possibly designed with an expected behavior when placed at a flow with a given velocity.
Interactions between a kirigami sheet and fluid flow constitute a multi-scale multi-physics FSI
problem. Since the flow that goes though the kirigami slits interacts locally with the strips, and this interaction influences the overall morphing behavior of the structure at the large scale, FSI must be considered at multiple
scales. Here we have shown how different types of responses of the structure and different fluid patterns in the wake of the structure can be obtained for different types of kirigami patterns. Some kirigami patterns deploy in a stable and progressive manner to large elongations when placed in fluid flow, while others undergo static instabilities and buckle at a critical flow velocity, or dynamic instabilities and exhibit limit cycle oscillations. These different types of structural responses result in a variety of fluid patterns in their wake. Small-scale vortex shedding, in sizes comparable with the size of the slits, is observed in the wake of a ribbon cut kirigami sheet when the structure only undergoes static deflections. These small-scale vortices give way to large-scale shedding of vortices in the wake, in sizes comparable with the size of the sheet, when the kirigami sheet undergoes limit cycle oscillations. The area behind the structure in a closed curve kirigami sheet experiences an almost still fluid condition despite the large velocity of the incoming flow, while in the case of fractal cuts, jets can be produced and directed in specific directions. 

In the future, Kirigami sheets can be designed such that their flow-induced instabilities or lack thereof correspond to a desired flow behavior in their wake. For example, the number of parallel vortices and the frequency of the vortices that are shed in the wake of a ribbon cut pattern, the size of the area of still fluid behind a closed curve kirigami pattern with a central disc, and the direction and strength of the jets that are formed in the wake of fractal-cut patterns can be manipulated by making changes in the designs of the kirigami sheets. Kirigami sheets can be designed such that at a specific flow velocity they assume a specific shape (such as the ones shown in Figure~\ref{fig:prel_DownStreamAndSideViews}) and therefore produce a desired wake. The ability to create controlled small-scale vortex shedding, induce desired flow-induced instabilities on structures, and form specifically-angled jets will enable several future applications in flow mixing (e.g., by producing small vortices in uniform flow at low Reynolds numbers), flow control (e.g., by controlling the direction and the number of jets that are produced downstream), and  underwater soft robotics (e.g., by imposing desired flow-induced oscillations on structures). 

\vspace{5pt}
\noindent
\textbf{Acknowledgements}

\noindent
This work was supported partially by the National Science Foundation, grant number CBET 2320300.\\

\bibliographystyle{elsarticle-harv} 
\bibliography{Main}

\end{document}